\documentclass[%
 aip,
 rsi,%
 amsmath,amssymb,
 preprint,%
]{revtex4-1}

\usepackage{graphicx}
\usepackage{dcolumn}
\usepackage{bm}
\usepackage[caption=false]{subfig} 
\usepackage{float}
\usepackage{multirow}

\begin{document}


\title{Widefield two-photon excitation without scanning: live cell microscopy with high time resolution and low photo-bleaching}

\author{Rumelo Amor}
 \email{rumelo.c.amor@strath.ac.uk}
\affiliation{ 
Centre for Biophotonics, Strathclyde Institute of Pharmacy and Biomedical Sciences, University of Strathclyde, 161 Cathedral Street, Glasgow G4 0RE, United Kingdom
}%

\author{Johanna Tr\"{a}g{\aa}rdh}
\affiliation{ 
Centre for Biophotonics, Strathclyde Institute of Pharmacy and Biomedical Sciences, University of Strathclyde, 161 Cathedral Street, Glasgow G4 0RE, United Kingdom
}%

\author{Gillian Robb}
\affiliation{ 
Centre for Biophotonics, Strathclyde Institute of Pharmacy and Biomedical Sciences, University of Strathclyde, 161 Cathedral Street, Glasgow G4 0RE, United Kingdom
}%

\author{Louise Wilson}
\affiliation{%
Neuroscience Group, Strathclyde Institute of Pharmacy and Biomedical Sciences, University of Strathclyde, 161 Cathedral Street, Glasgow G4 0RE, United Kingdom
}%

\author{Nor Zaihana Abdul Rahman}
\affiliation{%
Neuroscience Group, Strathclyde Institute of Pharmacy and Biomedical Sciences, University of Strathclyde, 161 Cathedral Street, Glasgow G4 0RE, United Kingdom
}%

\author{John Dempster}
\affiliation{%
Neuroscience Group, Strathclyde Institute of Pharmacy and Biomedical Sciences, University of Strathclyde, 161 Cathedral Street, Glasgow G4 0RE, United Kingdom
}%

\author{William Bradshaw Amos}
\affiliation{%
MRC Laboratory of Molecular Biology, Francis Crick Avenue, Cambridge Biomedical Campus, Cambridge CB2 2QH, United Kingdom
}%

\author{Trevor J. Bushell}
\affiliation{%
Neuroscience Group, Strathclyde Institute of Pharmacy and Biomedical Sciences, University of Strathclyde, 161 Cathedral Street, Glasgow G4 0RE, United Kingdom
}%

\author{Gail McConnell}
\affiliation{ 
Centre for Biophotonics, Strathclyde Institute of Pharmacy and Biomedical Sciences, University of Strathclyde, 161 Cathedral Street, Glasgow G4 0RE, United Kingdom
}%

\date{\today}

\begin{abstract}
We demonstrate fluorescence imaging by two-photon excitation without scanning in biological specimens as previously described by Hwang and co-workers, but with an increased field size and with framing rates of up to 100 Hz. During recordings of synaptically-driven Ca$^{2+}$ events in primary rat neurone cultures loaded with the fluorescent Ca$^{2+}$ indicator Fluo-4 AM, we have observed greatly reduced photo-bleaching in comparison with single-photon excitation. This method, which requires no costly additions to the microscope, promises to be useful for work where high time-resolution is required.
\end{abstract}

\maketitle


Immediately after its introduction by Denk, Strickler and Webb in 1990 \cite{Denk90}, two-photon excitation of fluorescence was adopted in many fields of biomedicine. Its chief advantages are its ability to penetrate more deeply into tissues than single-photon excitation, the creation of optical sections (by the combination of an excitation proportional to the square of the intensity with the conical beam geometry) and the more efficient utilization of scattered emission than is possible in a confocal microscope. The chief drawbacks were the slow scanning speed, which restricted the original instruments to a rate of approximately one image per second, often unusable for electrophysiology, where millisecond time resolution may be needed, and the high rate of photo-bleaching if the laser intensity was increased \cite{Patterson2000}. To overcome the slowness, faster scanning mirrors have been used \cite{Nguyen2001}, and parallelism has been achieved in two-photon imaging by the use of slit scanning \cite{Brakenhoff96} or the scanning of multiple foci, preferably uncorrelated in time \cite{Fittinghoff2000, Niesner2007, Deniset2007}.  Two-photon light-sheet microscopy has also been used to increase the rate of imaging \cite{Truong2011}, but because scattering and absorption in the specimen causes inhomogeneous illumination, this method is best suited to highly transparent tissue volumes. A simple and elegant approach of illumination with a large stationary spot of light from a mode-locked femtosecond-pulsed laser in a conventional microscope and imaging of the emission from structures within the spot, though not favoured by Fittinghoff \emph{et al.} who were apparently the first to try it \cite{Fittinghoff2000}, was shown by Hwang \emph{et al.} to provide images of good quality \cite{Hwang2011}. We show how this unexpected result can be explained by the vastly-increased exposure time of the simpler method.

The work of Hwang \emph{et al.} on several medically-important specimens seems to have only a limited effect on normal practice, possibly because the illuminated spot was no greater than 60 $\mu$m in diameter and the loss of optical sectioning has been regarded as serious. We have repeated and extended this approach, increasing the field size to a diameter of 90 $\mu$m and concentrating on fast transients in living neurones, a specimen where the potentially unlimited time resolution of the stationary beam can be evaluated. 

In this work, we have used modest average powers from an 80 MHz repetition rate femtosecond-pulsed Ti:Sapphire laser and have collected images using a sensitive sCMOS camera at frame rates up to 100 Hz. All components of our simple microscope setup are commercially available. We have explored the image quality, phototoxicity and bleach rate at different framing rates, making extended video recordings of synaptically-driven Ca$^{2+}$ events in live neurones loaded with Fluo-4 AM, which we have chosen because they show both slow and fast activity. Our chief new finding is that photo-bleaching is greatly reduced in the widefield two-photon regime relative to that seen with single-photon excitation. We have shown that this unexpected advantage relative to normal practice is particularly marked at high framing rates. 

\section*{Results}

\begin{figure}[H]
\centering
\includegraphics[width=\textwidth]{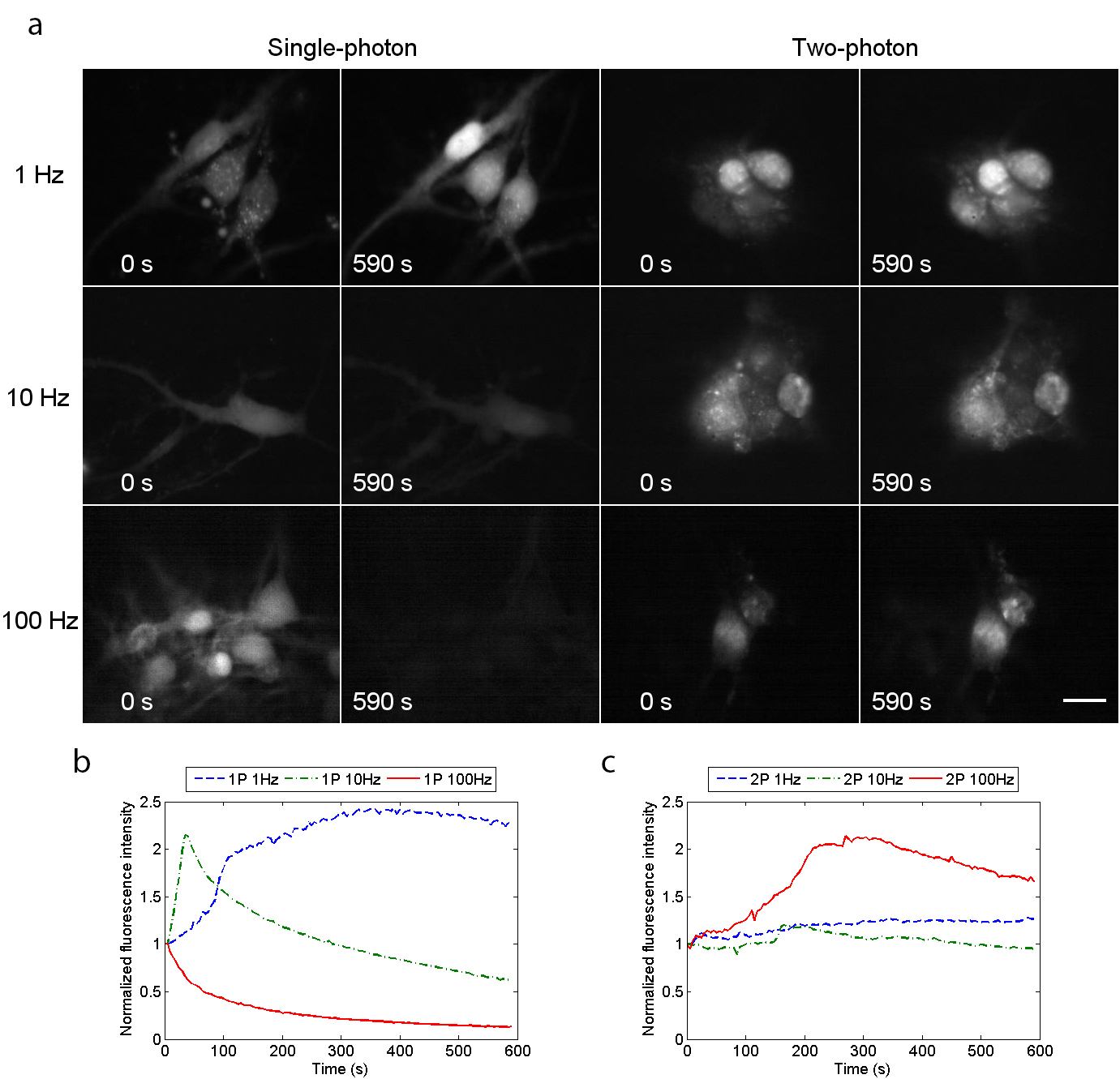}

\centering

\caption{Reduced photo-bleaching using widefield two-photon excitation. (\emph{a}) Single-photon and two-photon excited widefield images of live neuronal cells loaded with Fluo-4 AM, taken at frame rates of 1 Hz, 10 Hz and 100 Hz with continuous irradiation for 590 seconds. The normalised average fluorescence intensities are plotted versus time in (\emph{b}) for single-photon excitation and in (\emph{c}) for two-photon excitation. Photo-bleaching was observed when using single-photon excitation at 10 Hz and 100 Hz because of the higher light doses required to compensate for the short exposure times, whereas no photo-bleaching was observed when using two-photon excitation at similar frame rates. Scale bar=15 $\mu$m.}\label{Fig1} 
\end{figure}

We obtained high-resolution, high-contrast two-photon excited fluorescence images of Ca$^{2+}$ transients in live neuronal cell bodies at frame rates of up to 100 Hz. Figure~\ref{Fig1}a shows widefield single-photon and widefield two-photon excited fluorescence images of live cells with continuous irradiation. The frame rates were 1 Hz, 10 Hz and 100 Hz. The excitation power was adjusted to give a similar subjective appearance in the first images of all the series. This strategy was adopted to achieve similar signal-to-noise values at the start of all the series. The fluorescence signal intensities for three regions of interest (ROIs) located over neuronal cell bodies in each culture dish were measured from time t=0 to 590 seconds at 5-second intervals using ImageJ \cite{Schneider2012}. An average value of fluorescence intensity from the three ROIs was obtained for each dataset, and was then normalised relative to the fluorescence signal intensity at time t=0 seconds. The resultant plots of normalised average fluorescence signal intensity with time for single-photon excitation and two-photon excitation are presented as Figs.~\ref{Fig1}b and \ref{Fig1}c.

For both single-photon and two-photon excitation at a frame rate of 1 Hz, the overall fluorescence signal intensity increased over time but was much greater in the single-photon case, with an increase in signal in excess of 100 \% compared to a 25 \% increase using two-photon excitation over the same time period. However, at a frame rate of 10 Hz, there was very rapid fading using single-photon excitation (following a large Ca$^{2+}$ spike at around 30 seconds), whereas the fading using two-photon excitation was minimal, with less than 5 \% loss of fluorescence signal intensity after 590 seconds. At a frame rate of 100 Hz, however, recordings were quite different under the two illumination regimes. Using single-photon excitation, there was a rapid decrease in fluorescence signal intensity, with almost no fluorescence recorded after 590 seconds. For two-photon excitation, the fluorescence signal intensity increased slowly over the first 200 seconds and then began to decrease. So even at this fast framing rate (with increased irradiation to compensate for the shorter integration time), useful images were being obtained with widefield two-photon excitation long after the conventional single-photon recordings had faded totally. We noted also that that the rate of fluorescence intensity increase under two-photon excitation at 100 Hz was less rapid than that measured for single-photon excitation at the much slower speed of 1 Hz. 

Supplementary Video 1 shows time-lapse recordings of live neuronal cell bodies loaded with Fluo-4 AM at a frame rate of 100 Hz using single-photon (left) and two-photon (right) widefield excitation. The neuronal cultures were irradiated continuously for 590 seconds and the video shows a frame every 5 seconds, demonstrating very clearly the rapid decrease in fluorescence intensity observed using widefield single-photon excitation, while no significant reduction in fluorescence intensity was observed using widefield two-photon excitation. 
        
The mechanisms responsible for such light-induced changes over time are at present not understood. Increased fluorescence has often been observed and it has been suggested that this results from the photo-induced release of Ca$^{2+}$ \cite{Knight2003, McDonald2012} or from direct photo-activation of the fluorescent dye \cite{Betzig2006, Mitchison98} or effects on membrane permeability of sample heating \cite{Schonle98}, all of which can be expected at the short wavelengths applied in single-photon microscopy. The commoner problem of photo-bleaching due to the destruction of the dye by free-radicals generated by illumination \cite{Tsien2006} is a serious limitation in all single-photon live cell imaging. Our present results suggest that widefield two-photon microscopy may overcome the bleaching problem at frame rates between 10--100 Hz. 

\begin{figure}[H]
\centering
\includegraphics[width=0.6\textwidth]{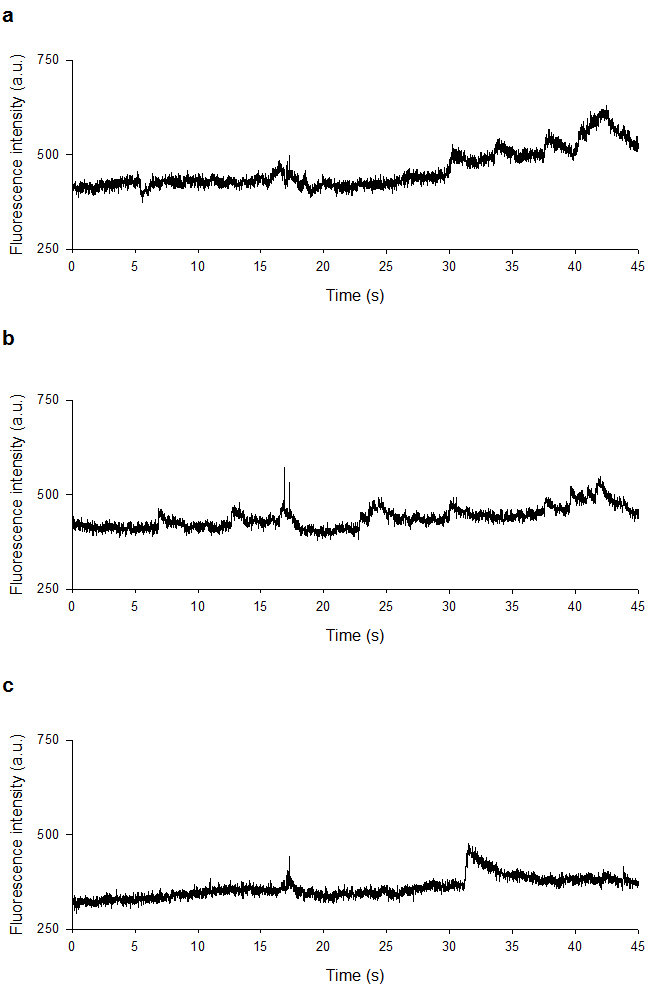}

\centering

\caption{Localisation of changes in fluorescence intensity. (\emph{a}), (\emph{b}) and (\emph{c}) show widefield two-photon excited fluorescence intensities within three adjacent live neuronal cell bodies loaded with Fluo-4 AM, acquired at a frame rate of 100 Hz over a duration of 45 seconds. This confirms that the measured change in fluorescence signal intensity with time was not global across the image, but was instead localised to individual cell bodies at different times.}\label{Fig2} 
\end{figure}

To confirm the two-photon nature of the excitation process, we obtained images with frame rates of 1 Hz, 10 Hz, 50 Hz and 100 Hz at five time-averaged incident powers. Log-log plots of fluorescence intensity from five ROIs with excitation power gave gradients of between 1.8 and 2.3 for all frame rates, confirming two-photon excitation. An example time course with three ROIs is shown in Fig.~\ref{Fig2}, and Supplementary Video 2 shows slow ($\sim$seconds) Ca$^{2+}$ transients at a camera frame rate of 100 Hz over a duration of 10 seconds. For display purposes, the video file has been reduced in size by 50 \%. In this video, three ROIs were chosen in three adjacent cell bodies to not only demonstrate the rate of image capture but also to confirm that the measured change in fluorescence signal intensity with time was not global across the image, but is localised to individual cell bodies at different times. This evidence suggested that the change in measured fluorescence signal intensity was not a consequence of fluctuations in laser power or camera instability, but arose from localised changes in intracellular Ca$^{2+}$ concentration.

\begin{figure}[H]
\centering
\includegraphics[width=0.6\textwidth]{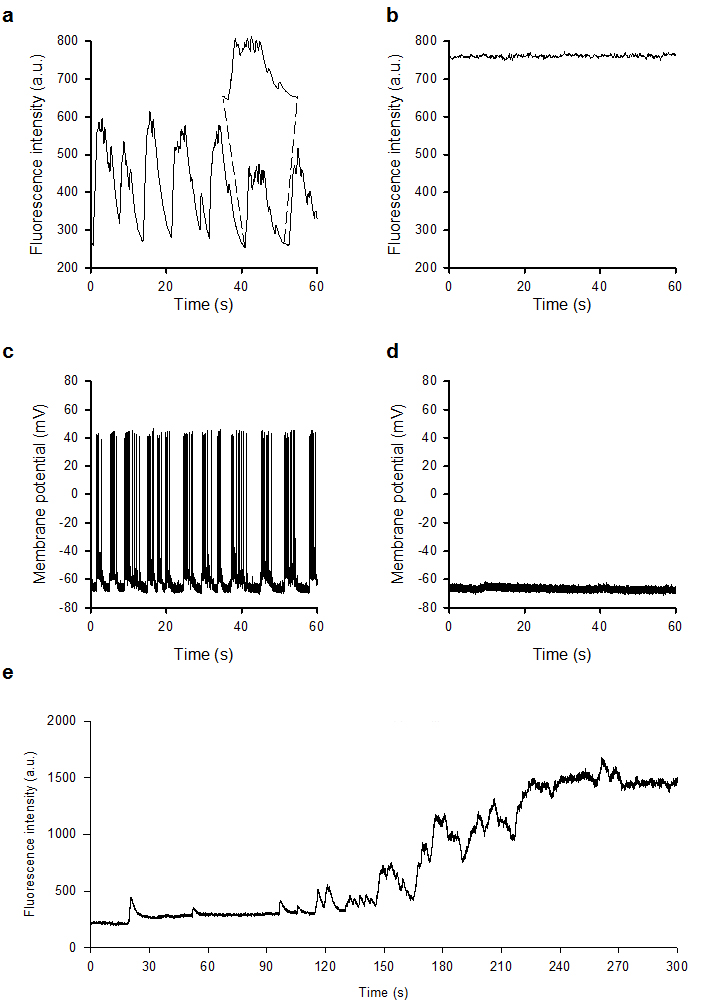}

\centering

\caption{Widefield two-photon excitation for recording fast synaptic activity. (\emph{a}) Fluorescence recording at a frame rate of 10 Hz shows spontaneous changes in fluorescence intensity in live neuronal cell bodies loaded with Fluo-4 AM, which were abolished in the presence of the glutamatergic antagonists DL-AP5 and NBQX (\emph{b}). The expanded region of (\emph{a}) shows the envelope of events indicating fast activity. (\emph{c}) Whole-cell current clamp recordings from hippocampal neurones revealing spontaneous action potential firing which is abolished in the presence of DL-AP5 and NBQX (\emph{d}). (\emph{e}) Fluorescence recording at 50 Hz following the application of bicuculline to show the consequence of reducing the inhibitory tone present in the cultures. Spontaneous changes in fluorescence intensity initially became more frequent and erratic in nature but at later time points were not present given the large rise in fluorescence within the neuronal cells due to uncontrolled glutamatergic activity.}\label{Fig3}
\end{figure}

Having established that wide-field two-photon excitation reduces photo-bleaching at high frame rates, we then examined whether Ca$^{2+}$ events induced by synaptic activity could be observed in primary hippocampal cultures. Figure~\ref{Fig3}a and Supplementary Video 3 (left panel), acquired at a frame rate of 10 Hz, show spontaneous changes in fluorescence intensity in live neuronal cell bodies over a 60-second time period, which were abolished in the presence of the AMPA and NMDA receptor antagonists NBQX (20 $\mu$M) and DL-AP5 (100 $\mu$M) (Fig.~\ref{Fig3}b and right panel of Supplementary Video 3), indicating that these events are driven by glutamatergic excitatory synaptic activity. The sensitivity to NBQX/DL-AP5 is similar to that observed when using the whole-cell patch clamp technique to monitor synaptically driven events including spontaneous action potential firing using identical cultures (Fig.~\ref{Fig3}c and \ref{Fig3}d) \cite{Gan2011, Ledgerwood2011} thus highlighting the functional capability of this method. 

Additionally, in separate experiments, we show the consequence of reducing the intrinsic inhibitory tone present in the cultures through the application of the GABA$_\mathrm{A}$ receptor antagonist bicuculline (20 $\mu$M). Figure~\ref{Fig3}e and Supplementary Video 4, acquired at a frame rate of 50 Hz, show that spontaneous changes in fluorescence intensity initially became more frequent and erratic in nature but at later time points were not present given the large rise in fluorescence within the neuronal cell bodies due to uncontrolled glutamatergic activity. 

\begin{figure}[H]
\centering
\includegraphics[width=\textwidth]{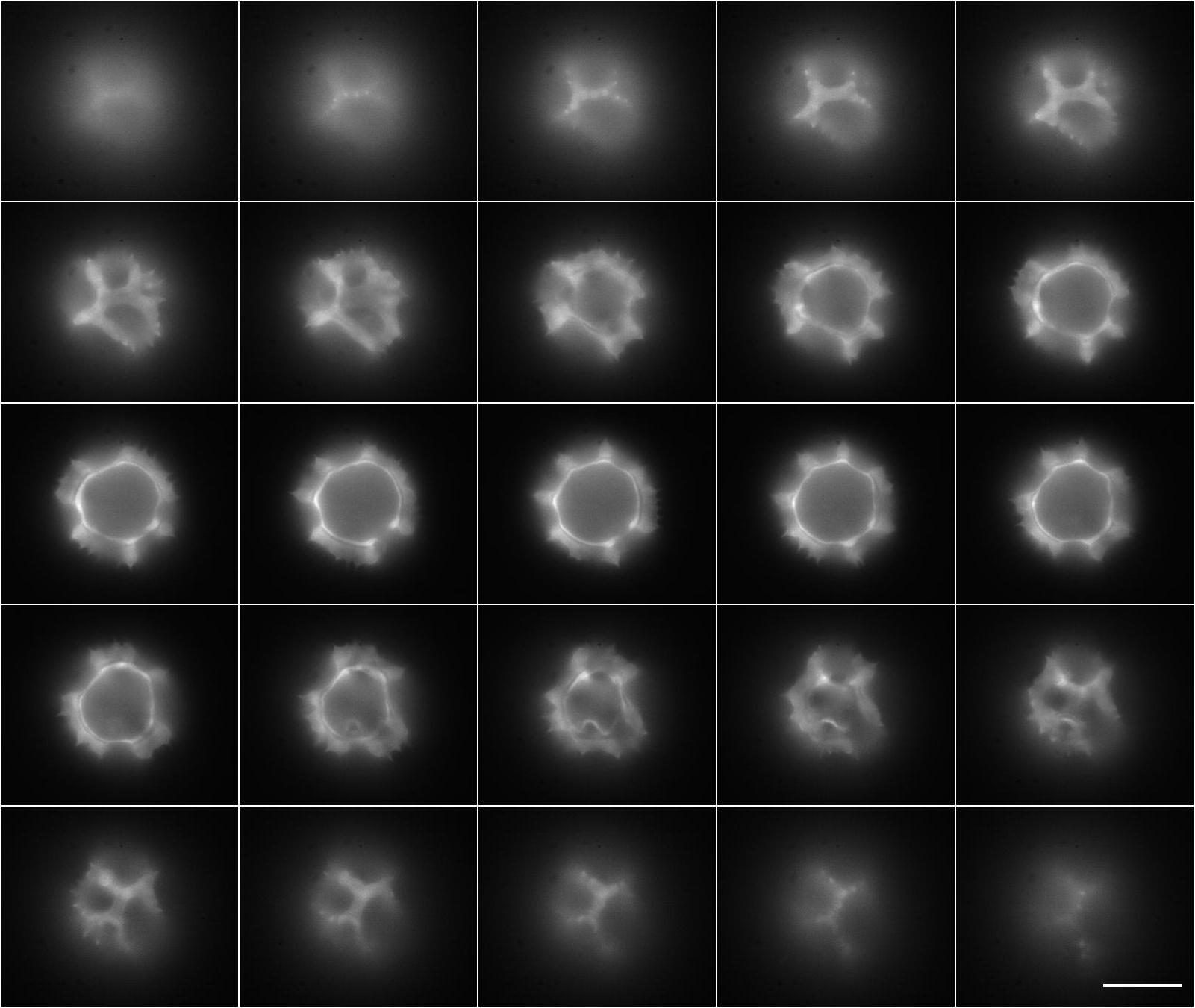}

\centering

\caption{Optical sectioning of pollen specimen. Widefield two-photon microscopy shows some optical sectioning of an auto-fluorescent fixed \emph{Taraxacum} pollen specimen, obtained by moving the specimen by 1 $\mu$m increments axially over a range of 25 $\mu$m. No post-processing was performed on the images except for cropping to display only a single pollen grain within the image field. The optical sectioning shown here is closely similar to that obtained in a widefield single-photon fluorescence microscope (results not shown). This confirms that the optical depth of field is due to the focusing of the emission only. Scale bar=15 $\mu$m.}\label{Fig4}
\end{figure}

\begin{figure}[H]
\centering
\includegraphics[width=0.85\textwidth]{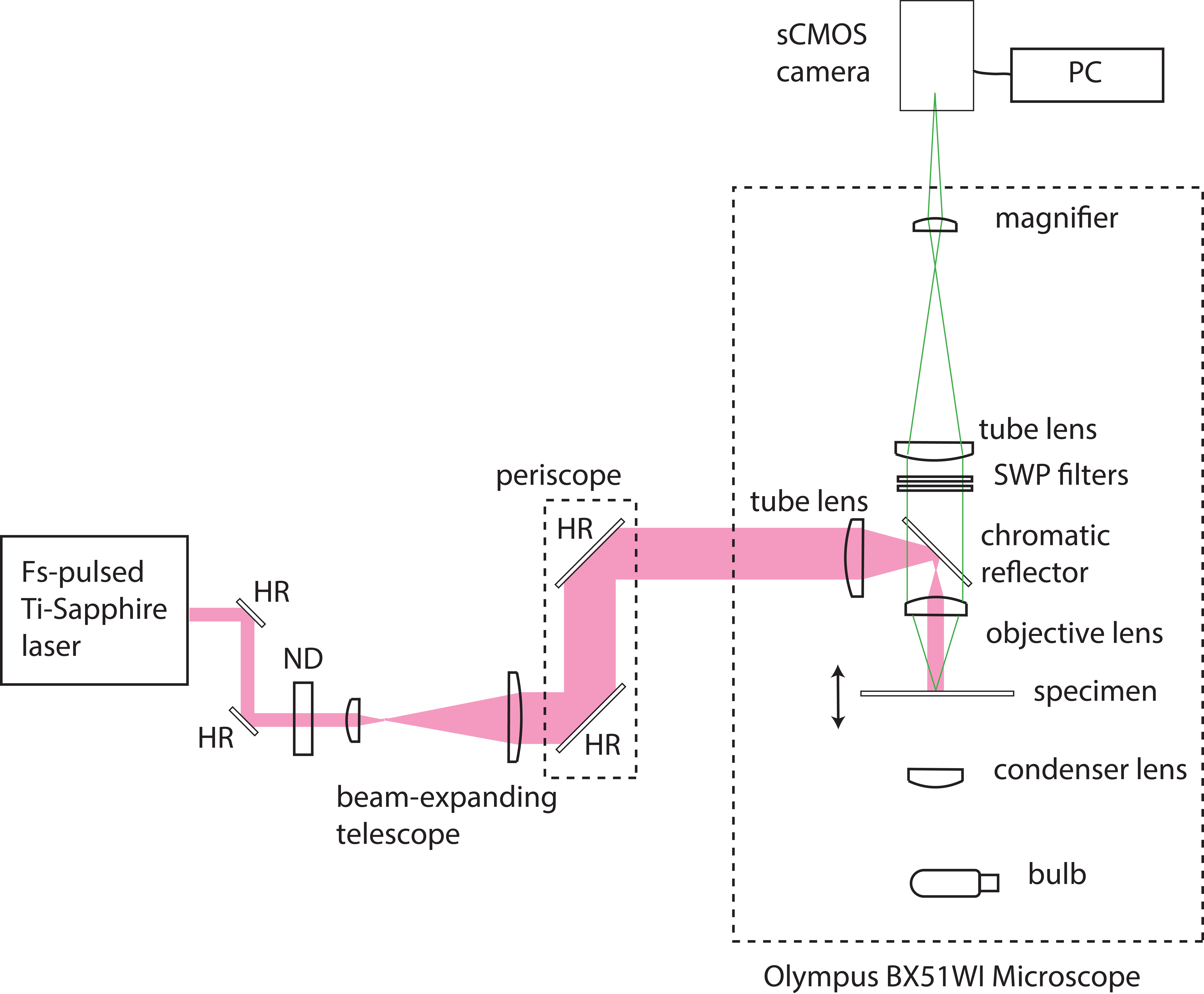}

\centering
\caption{Widefield two-photon microscope and image capture system. An upright microscope was modified for two-photon excitation by changing the excitation, emission and dichroic filters. A cooled sCMOS camera controlled by a PC running the freely available software WinFluor was used to detect fluorescence from the specimen. A wavelength-tunable femtosecond-pulsed Ti:Sapphire laser was used as the excitation source. The weakly divergent output from the Ti:Sapphire laser was coupled into the microscope, and focused to provide a beam waist close to the back aperture of the objective lens. This produced a wide and weakly-focused beam in the specimen plane which did not contribute significantly to the optical sectioning power. HR=highly reflecting mirror, ND=neutral density filter, SWP=short-wave pass filter.}\label{Fig5} 
\end{figure}

Since the cell preparations used here were a thin monolayer of cells, we did not perform optical sectioning. However, some optical sectioning of thicker specimens was possible. Figure~\ref{Fig4} shows optical sectioning of an auto-fluorescent fixed \emph{Taraxacum} pollen specimen mounted in Histomount, obtained at an excitation wavelength of $\lambda$=820 nm and using the same chromatic reflectors and emission filters shown in Fig.~\ref{Fig5}. Here a 60x/1.35 NA oil immersion objective was used, and the montage was obtained by moving the specimen by 1 $\mu$m increments axially over a range of 25 $\mu$m. The frame rate was set to 10 Hz for each image. The spikes at the top and bottom of the pollen grain are clearly independently resolved without deconvolution or other image processing methods.

\section*{Discussion}

In evaluating this widefield two-photon microscopy method, it is natural to ask what advantages it confers in relation to conventional two-photon scanning microscopy.

The most important advantage of the method is the low photo-bleaching rate, compared with both two-photon scanning and single-photon widefield methods, when applied to the imaging of fast events. This would appear to present a great experimental advantage. The reason for this advantage is almost certainly that our widefield regime requires an excitation intensity that is about three orders of magnitude less than that used in point-scanning two-photon microscopy. Lowering the excitation intensity may bring a disproportionate benefit, since it is known that photo-bleaching has a dependency on intensity even higher than two-photon fluorescence \cite{Patterson2000, Patterson97}. We suggest that our method may prove advantageous when using fluorophores or photoproteins with fast photo-bleaching rates, including Ca$^{2+}$-sensitive dyes, CFP \cite{Malkani2011}, eBFP \cite{Zaccolo2000} and YFP \cite{McAnaney2005}. This technique may also reduce unwanted photo-bleaching in FRAP, FLAP and other similar experimental methods, where photo-bleaching of the untargeted region can compromise results \cite{Weiss2004, Dunn2002}. We have used a single camera and fluorophore here, but our method could be easily adapted for simultaneous multi-channel recording using an image splitter and a second sCMOS camera.

The advantage in time-resolution is plainly shown by the results at high frame rates. The penetration ability of two-photon, as compared with single-photon excitation, which has been well established \cite{Centonze98}, has not been measured here, but is expected in this method also, because it depends on the lower scattering of longer wavelengths and the low absorption of the ultra-short pulsed infra-red excitation wavelengths. The widefield two-photon method cannot be expected to perform as well as conventional scanned two-photon excitation in optical sectioning ability. By imaging fluorescent test specimens using the 60x/1.35 NA oil immersion lens at $\lambda$=820 nm, we measured an axial point spread function of around 5 $\mu$m and a measured lateral resolution of 850 nm. These resolution values are lower than for a point-scanning two-photon microscope system with similar excitation parameters, but they are an expected consequence of weak focusing of the excitation beam. Nevertheless, the results from our live cell experiments and imaging of thicker specimens suggest that this method could be adapted for in vivo imaging, with the advantage of reduced photo-bleaching. Although no temporal focusing \cite{Oron2005} was used here, our widefield method could be used with temporal focusing to obtain improved optical sectioning, without the low image contrast and much reduced resolution observed in multi-focal two-photon microscopy \cite{Bewersdorf98}.

Ultra-short pulsed near-infrared lasers of the type used here have been previously used to generate Ca$^{2+}$ waves in differentiated cells. Smith \emph{et al.} \cite{Smith2006} used a single diffraction-limited beam focus of $w_0$=0.3 $\mu$m, with an average power $P_{av}$ \textgreater~20 mW, $\lambda$=775 nm, repetition frequency $\Delta{\nu}$=82 MHz and pulse duration $\tau$=140 fs, which gives a peak intensity $I_{peak}$ \textgreater~6.16 $\times$ 10$^{15}$ W/m$^2$.  However, since the peak intensity of illumination used in our two-photon widefield microscope is around three orders of magnitude less than Smith \emph{et al.} used in their experiments, it is highly unlikely that we are observing light-induced Ca$^{2+}$ transients. Indeed, the sensitivity of the observed Ca$^{2+}$ events to blockade of glutamatergic synaptic activity indicates that the events are synaptically driven and are presumably triggered by action potentials arriving at the neuronal cell body, the consequence of which is the transient increase in intracellular Ca$^{2+}$ level due to membrane potential depolarisation. The close similarity between the optical transients observed here and those recorded electrically in individual cells by the whole-cell path clamp technique in current clamp mode are a clear demonstration of the sensitivity, time resolution and usefulness of this method. We have concentrated here on the improved time resolution and remarkably low level of photo-bleaching, but, as already pointed out by Hwang \emph{et al.}, there may be other advantages over single-photon imaging, such as improved discrimination against auto-fluorescence and clearer imaging of dense tissues \cite{Hwang2011}.

\section*{Methods}

\textbf{Excitation and detection of two-photon fluorescence.} Awareness of the high peak intensities required for two-photon excitation has probably deterred experimentation on widefield two-photon microscopy. Here we compare excitation and detection parameters in the widefield two-photon microscope with that in a standard point-scanning two-photon microscope.\\

For the laser source, we assume a wavelength of $\lambda$=780 nm, repetition frequency $\Delta{\nu}$=80 MHz and pulse duration $\tau$=140 fs with a 60x/0.9 NA water dipping lens, as used in our live cell imaging experiments. We also assume that pulse stretching is negligible, and is similar in both the laser scanning and widefield microscopes.\\

In a point scanning two-photon microscope, we can assume a time-averaged power at the specimen plane of $P_{av}$=15 mW. With a pulse duration $\tau$ and pulse repetition frequency $\Delta{\nu}$, from
\begin{equation}
P_{peak}=\frac{P_{av}}{\tau\Delta\nu}
\label{Eq1}
\end{equation}
a peak power of $P_{peak}$=1.34 kW is obtained.\\

The beam waist radius ($w_0$) of the excitation laser at the specimen plane is given by \cite{Kogelnik66}
\begin{equation}
w_0=\frac{\lambda}{\pi\theta},
\end{equation}
where $\theta$ is the half-angle of the beam divergence given by the numerical aperture
\begin{equation}
NA=nsin{\theta}.
\end{equation}
For the 60x/0.9 NA water dipping lens ($n$=1.33), $\theta$=0.74 rad and thus at a wavelength of $\lambda$=780 nm, the beam waist radius is calculated to be $w_0$=334 nm.\\

The peak intensity in the excitation spot is given by
\begin{equation}
I_{peak}=\frac{P_{peak}}{\pi{w_0}^2},
\label{Eq4}
\end{equation}
and from the calculated values of $P_{peak}$ and $w_0$, a peak intensity of $I_{peak}$=3.82 $\times$ 10$^{15}$ W/m$^2$ is obtained for point-scanning.\\

In the widefield two-photon microscope, we use a time-averaged power of $P_{av}$=360 mW at the specimen plane which, from Eq.~\ref{Eq1} when considering the same pulse duration and pulse repetition frequency, gives a peak power of $P_{peak}$=32 kW.\\

Instead of focusing to a diffraction limited spot using the same 60x/0.9 NA lens by overfilling the back aperture, we instead consider focusing to a small spot near the back aperture to obtain a weakly focusing spot at the specimen plane with a radius of $w_0$=45 $\mu$m. Using these values of $P_{peak}$ and $w_0$ in Eq.~\ref{Eq4} above, the peak intensity in the widefield two-photon microscope is therefore $I_{peak}$=5.03 $\times$ 10$^{12}$ W/m$^2$.  We note that this peak intensity is approximately three orders of magnitude lower than the peak intensity for a point-scanning two-photon microscope.\\

The substantially lower peak excitation intensity in the widefield two-photon microscope is compensated by negating the need to scan the beam. By continuously irradiating across the whole field, we can use a much longer integration time for collection of fluorescence.\\

In a point-scanning two-photon microscope, we can assume a point dwell time of T=5 $\mu$s and, since the fluorescence lifetime of the fluorophore is very short (in the order of ns), we can assume that fluorescence is generated and collected over the same 5 $\mu$s duration.  For an image that is 640 $\times$ 560 pixels, which is the same image size we use in the widefield setup, this point dwell time yields a frame rate of 0.56 Hz. The excitation radiation applied to the specimen for each point is therefore $I_{peak}$ $\times$ T=1.91 $\times$ 10$^{10}$ W$\cdot$s/m$^2$.\\

In the widefield two-photon microscope, the specimen is continuously irradiated with the femtosecond-pulsed laser radiation, but in comparison with the point-scanning two-photon microscope we can consider the dwell time to be equivalent to the frame rate of the camera.  For images taken at 100 Hz (i.e. a T=10 ms exposure time), the excitation radiation applied to the specimen during the image capture can be considered to be $I_{peak}$ $\times$ T=5.03 $\times$ 10$^{10}$ W$\cdot$s/m$^2$. The overall total excitation radiation applied in the widefield two-photon microscope is therefore similar to that used in point-scanning two-photon microscopy.\\

In summary, when using widefield excitation, the peak intensity is orders of magnitude lower than in point-scanning excitation and the total excitation radiation applied is similar. By collecting the light on all the pixels in parallel we achieve long per pixel integration times, and can collect enough photons for imaging at fast frame rates. Moreover, the quantum efficiency of an sCMOS camera (60 \% for the Zyla camera) is likely to be higher than the efficiency of a photomultiplier tube used in a point-scanning two-photon microscope ($\sim$30 \%), therefore offering the possibility of high detection sensitivity.\\

\textbf{Experimental setup.} Our simple widefield two-photon microscope was configured as shown in Fig.~\ref{Fig5}. A commercial upright epi-fluorescence microscope (BX51WI, Olympus) was modified for two-photon excitation. A chromatic reflector (FF670-SDi01, Semrock) reflecting wavelengths longer than 670 nm and transmitting shorter wavelengths was used to direct excitation towards the specimen and to transmit fluorescence. A cooled sCMOS camera (Zyla 5.5, Andor) controlled by a PC running the freely available software WinFluor \cite{Dempster2014} was used to detect fluorescence from the specimen. For blocking of the laser source from the camera and to ensure that only the fluorescence signal contributed to the image, we used 680 nm and 694 nm short-wave pass filters (FF01-680/SP and FF01-694/SP, Semrock) in the collection path. A commercially available wavelength-tunable femtosecond-pulsed Ti:Sapphire laser (Chameleon Ultra II, Coherent) was used as the excitation source. This laser delivered a maximum time-averaged power of 2.3 W at a repetition rate of 80 MHz with a pulse duration of 140 fs. The weakly divergent output from the Ti:Sapphire laser was steered using a pair of highly reflecting mirrors (BB1-E03, Thorlabs) and then attenuated using a variable neutral density filter (NDC-25C-4, Thorlabs) and expanded to a diameter of 22 mm using a plano-convex lens pair of focal lengths f=+35 mm and f=+100 mm (LA1027-B and LA1509-B, Thorlabs). The beam height was changed using a two-mirror periscope (BB1-E03, Thorlabs) to permit coupling of the laser into the upright microscope. At height, the beam was focused using a single plano-convex lens of focal length f=+75 mm (LA1608-B, Thorlabs), with the beam waist close to the back aperture of the objective lens. This produced a wide and weakly-focused beam in the specimen plane which did not contribute significantly to the optical sectioning power. A beam diameter was found experimentally which allowed excitation of the dye but with minimal photo-bleaching at a time-averaged laser power of 360 mW at a wavelength of 780 nm. The illuminated area then had a diameter of 90 $\mu$m and a 1.6$\times$ magnifier was used to fill the camera sensor. All live cell imaging data presented here used a 60x/0.9 NA water dipping lens (Olympus).\\

For single-photon widefield imaging experiments, the short-wave pass filters were removed and the long-wavelength chromatic reflector was replaced with a filter cube (XF22, Omega Optical, Inc.) that comprised an excitation filter with a peak wavelength of 485 $\pm$ 22 nm, a chromatic reflector with a 505 nm long-wave pass filter and an emission filter with a peak wavelength of 530 $\pm$ 30 nm. The 75 mm plano-convex tube lens was removed, and the mercury arc lamp excitation source (U-ULS100HG, Olympus) and coupling optics originally supplied with the microscope were reattached. At a wavelength of 485 nm, the maximum time-averaged power at the specimen plane exiting the 60x/0.9 NA water dipping objective lens was 240 $\mu$W. The same camera was used with the same PC and controlling software as with widefield two-photon excitation.\\

\textbf{Imaging primary rat hippocampal cultures.} We performed widefield two-photon microscopy of primary rat hippocampal cultures to demonstrate the capability of rapid imaging. Primary rat hippocampal cultures were prepared as described in Ref. 17. Briefly, Sprague-Dawley rat pups 1--2 days old were killed by cervical dislocation and decapitation, following UK Home Office guidelines, and the brain removed. The hippocampi were then dissected out, incubated in a papain solution (1.5 mg/ml, Sigma-Aldrich) at 37 $^\circ$C for 20 minutes. The hippocampi were then washed in solution containing bovine serum albumin (10 mg/ml), dissociated by trituration and plated onto coverslips previously coated with poly-L-lysine (0.1 mg/ml) at a final density of 3 $\times$ 10$^5$ cells/ml. Cultures were incubated in Neurobasal-A Medium (Invitrogen) supplemented with 2 \% (v/v) B-27 (Invitrogen) and 2 mM L-glutamine and maintained in a humidified atmosphere at 37 $^\circ$C/5 \% CO$_2$. After 5 days \emph{in vitro} (DIV), cytosine-D-arabinofuranoside (10 $\mu$M) was added to inhibit glial cell proliferation. Cells were used experimentally from 11-14 DIV. \\

For imaging, hippocampal cultures were washed twice with a HEPES-buffered saline (HBS) containing (in mM): NaCl 140, KCl 5, MgCl$_2$ 2, HEPES 10, D-glucose 10 and CaCl$_2$ 2, pH 7.4, and transferred to HBS containing Fluo-4 AM (10 $\mu$M, 45--60 min, room temperature). Once loaded, cells were washed with HBS and images were obtained from the cell bodies of neurones with constant irradiation and with frame rates of up to 100 Hz. Pixel binning was used to increase the signal to noise ratio of the images: four pixel binning was used at frame rates of 1--50 Hz, and eight pixel binning was used for camera frame rates of 100 Hz. Experiments were performed on cultures at room temperature. Cells were identiﬁed as the cell bodies of neurones based on their morphological characteristics. Data were calculated as changes in fluorescence ratio from the fluorescence signal intensity at time t=0 s.\\

We also compared the measured fluorescence signal intensity over time in the live cell neuronal preparations using both single-photon and two-photon widefield excitation to investigate photo-bleaching at different frame rates. For these measurements, we used time-averaged powers of $\sim$50 $\mu$W for single-photon and $\sim$100 mW for two-photon excitation at 1 Hz, $\sim$400 $\mu$W and $\sim$170 mW at 10 Hz, and $\sim$2.5 mW and 360 mW at 100 Hz. These time-averaged powers were chosen to facilitate the best comparison of single-photon and two-photon excitation. Similar initial fluorescence signal intensity counts were obtained by adjusting the excitation radiation in one region of the specimen then moving to an adjacent region for recording. It is perhaps helpful to note here that this setup procedure was possible by eye, since the two-photon excitation of fluorescence could be observed using the binocular viewer of the microscope. To our knowledge, this is the first observation of this kind.\\

To determine whether the transient changes in fluorescence signal intensity over time were caused by synaptically-driven events, NBQX (20 $\mu$M, Abcam) and DL-AP5 (100 $\mu$M, Abcam), AMPA receptor and NMDA receptor antagonists respectively, were applied in order to block excitatory synaptic activity. Additionally in separate experiments, the GABA$_\mathrm{A}$ receptor antagonist bicuculline methiodide (20 $\mu$M, Abcam), was applied to determine the consequence of reducing the intrinsic inhibitory tone present in the cultures. Data were recorded using widefield two-photon excitation ($\lambda$=780 nm, $P_{av}$=180 mW and a frame rate of 10 Hz for NBQX/DL-AP5, $P_{av}$=250 mW and a frame rate of 50 Hz for bicuculline) before either the NBQX/DL-AP5 or bicuculline were added to the bath. The excitation source was then blocked to prevent irradiation of the specimen for 10 minutes, and then further data were recorded using the same excitation and detection parameters.\\

\subsection*{Acknowledgements}

This research was supported in part by the Medical Research Council (grant no. MR/K015583/1) and the Engineering and Physical Sciences Research Council (grant no. EP/I006826/1). R.A. is supported by a Scottish Universities Physics Alliance INSPIRE studentship.

\subsection*{Author contributions}

W.B.A., T.B. and G.M. designed the study. J.D. developed the image acquisition and analysis program WinFluor. J.T. performed calculations. L.W. and N.Z.A.R. prepared primary rat hippocampal cultures. W.B.A. and G.M. prepared the pollen specimen. R.A., G.R. and G.M. performed the experiments. R.A., T.B., W.B.A. and G.M. analyzed the data and wrote the manuscript. 

\subsection*{Competing financial interests}

The authors declare no competing financial interests.

\newpage

\section*{Figures}

\end{document}